\documentclass[twocolumn,preprintnumbers,amsmath,amssymb]{revtex4}
\usepackage{graphicx}
\usepackage{dcolumn}
\usepackage{bm}
\usepackage{amsmath}
\usepackage{amssymb}
\usepackage{amsthm}
\usepackage{amsfonts}
\usepackage{enumerate}
\usepackage{latexsym}
\usepackage{hyperref}
\usepackage{centernot}
\begin{document}

\title{Seiberg-Witten monopoles: Weyl metal coupled to chiral magnets}

\author{Yue Yu}

\affiliation {Department of Physics, Center for Field
Theory and Particle Physics, State Key Laboratory of Surface Physics and Collaborative Innovation Center of
 Advanced Microstructures, Fudan University, Shanghai 200433, China}

\begin{abstract}
We study a Weyl (semi)metal which couples to local magnets. In the continuum limit, the Hamiltonian of the system matches the Chern-Simons-Maxwell-Dirac functional and then the ground state  is governed by  generalized Seiberg-Witten (SW) or Freund  equations in terms of the sign of Dzyaloshinskii-Moriya coupling.  The ground states determined by the Freund equations may either be monopolar Weyl semimetal accompanied by the ferromagnetic magnets (MWFM) or SW monopoles which consist of spheric Weyl fermions coupled to chiral magnets, depending on the strength of the Kondo coupling. In the latter phase, the topological ground state is characterized by SW invariants and with a Weyl surface on which the Weyl metal is of an exotic dispersion $\propto \sqrt k$. There are also the metastable SW monopole solutions carrying an opposite SW invariant for the SW equations while the ground state in this case is the MWFM state.
\end{abstract}


\date{\today}

\maketitle

In 1994, Seiberg and Witten (SW) found that the topological strong coupling supersymmetric Yang-Mill theory with $SU(2)$ instantons can be dual to a weak coupling $U(1)$  supersymmertic Yang-Mill theory, which was called S-duality \cite{SW}.  This breakthrough opened the prelude to the second superstring revolution.  Witten pointed out that associated with SW monopole equations (SWMEs) there is a topological invariant  which is  equivalent to the Donaldson invariant in the strong coupling theory \cite{ds2} but the former is easier to be calculated due to the abelian nature of the gauge group \cite{witten}.  This  developed a new area in mathematics \cite{donaldson}. 

On the other track, researches to topological states of matter have become the main theme in  fundamental condensed matter  physics during the past decade.  Besides the classical topological number, TKNN Chern number, in quantum Hall effects \cite{TKNN},  a full  classification of topological insulators and topological superconductors \cite{class1,class2} was set up after the discovery of the celebrated Kane-Mele $Z_2$ invariant \cite{KM}. A corresponding classification of topological metal and semimetal was also found \cite{vol,class6,class3,class4,class5,class7}. On the material side, large classes of two and three dimensional topological insulators as well as gapless topological states, such as Dirac and Weyl semimetals, were predicted and discovered \cite{KZ,QZ,BH,d1,d2,d3,d4,d5,w1,w2,w3,w4,w5,w6,w7}.

Comparing with the aforesaid progresses in the "topology of the band theory", studies on the topological nature of physical objects possess a much longer history and much wider area spectra after Dirac magnetic monopole was proposed \cite{dirac}. Among numerous instances, we mention the non-collinear and non-coplanar spin texture configurations in noncentrosymmetric systems, which are relevant to our study in this Letter. The competition between the Heisenberg exchange interaction and the Dzyaloshinskii-Moriya (DM) interaction $D$ \cite{Dz,Mo} may course fruitful spin textures such as helical/conical  spin structure and skyrmion configuration \cite{sky1,sky2}. We will see that the SW monopoles, identified as spheric Weyl fermions coupled to chiral magnets, can emerge in systems in which both the local magnets and Weyl fermions exist. Here we dub the "spheric Weyl fermions" because the Weyl fermions are massless on a wave vector sphere $S^2$, the "Weyl surface". On the Wely surface, the dispersion is exotic, i.e., $\propto\sqrt k$.  For $D>0$, the SW monopoles appear in the ground states of the system. There is a phase transition from the SW monopoles to the monopolar Weyl fermions coupled to a ferromagnetic magnetic order (MWFM) in a quantum critical point which is determined by the Kondo coupling strength $|K|$ between the Weyl fermion and local magnetization. These SW monopoles are the solutions of a variant of SWMEs, Freund equations (F-Eqs) \cite{freund}, and carry the SW invariant $SWI=+1$. For $D<0$, the SW monopoles with $SWI=-1$, corresponding to a generalization of the original SW equations (SW-Eqs)\cite{SW}, are metastable as they are the solutions of the SW-Eqs but are not stable against the MWFM state.  We expect the SW monopoles can emerge in existed noncentrosymmetric materials with ferromagnetic magnets.

We first give a synopsis of SW monopoles on a flat three-dimensional space $X$ \cite{suppl,2d}.    
 The SWMEs can be obtained by minimizing the Chern-Simons-Dirac functional on $X$ which reads  \cite{donaldson} 
  \begin{eqnarray}
\int d^3{\bf r}[ -i\chi^\dag \boldsymbol\sigma\cdot(\nabla+i{\bf A})\chi\pm\epsilon_{abc}A_a\partial_b A_c]~\label{CSD}
\end{eqnarray}
where $\chi^\dag=(\alpha^*,\beta^*)$ is a Weyl spinor; $A_a$ is a $U(1)$ gauge field; $\boldsymbol\sigma$ are Pauli matrices.  The repeat indices  imply summation over $a,b,c=1,2,3$.  By variating with $\chi^\dag$ and $A_a$, one has the SWMEs 
\begin{eqnarray}
\boldsymbol\sigma\cdot(\nabla+i{\bf A})\chi=0,~~\chi^\dag\sigma^a\chi=\pm\epsilon_{abc}\partial_bA_c. \label{sw2}
\end{eqnarray}
The equations with the plus sign in (\ref{sw2}) are called SW-Eqs while the minus sign corresponding to F-Eqs \cite{freund,freund1}.  The solution of SWMEs, a pair of $(A_a,\chi)$,  is called {\it SW monopole} \cite{SW,witten}. The square integrable (or periodic) solution space of the SWMEs is named as the moduli space of the SW monopoles \cite{witten,hut}. The SW invariant is basically defined by the Euler number of the moduli space  with its orientation\cite{witten,donaldson,bohn} and is equivalent to Casson invariant \cite{fl,donaldson}.

To solve Eqs.  (\ref{sw2}) on $X$, one can transfer the SWMEs to an ansatz \cite{sol1,sol2,sol3}
\begin{eqnarray}
&&\chi=\frac{1}{\sqrt{2(B_0+B_{03})}}\left(\begin{array}{c} B_0+B_{03}\\B_{01}+iB_{02}\end{array}\right) \label{4}\\
&&{\bf A}=-\frac{\nabla\times {\bf B}_0}{2B_0}-\frac{B_{01}\nabla B_{02}-B_{02}\nabla B_{01}}{2B_0(B_0+B_{03})} \label{5} \\
&&{\bf B}_0=\pm{\bf B}=\pm\nabla\times {\bf A} \label{6}
\end{eqnarray} 
with ${\bf B}_0=\chi^\dag\boldsymbol\sigma\chi$  obeying $\nabla\cdot {\bf B}_0=0$. One can check $\chi^\dag\chi=B_0$.  Many solutions of the SWMEs on  $R^3$ were known \cite{freund,freund1,sol1,sol2,sol3}.  
   
\noindent{\it Physical model. } We now study a physical system which consists of Weyl (semi)metal coupling to a local magnets. The system can be approximately described  by a continuum model Hamiltonian on $T^3$, 
\begin{eqnarray}
H&=&\int_{T^3}d^3{\bf r}\sum_i\biggl[\mp i\psi^\dag _{i,\pm}\boldsymbol\sigma\cdot\nabla\psi_{i,\pm}+K{\bf M}_i\cdot{\bf s}_{i,\pm}\nonumber\\
&+&\frac{D}4{\bf M}_i\cdot \nabla\times{\bf M}_i+\frac{J}2(\nabla {\bf M}_i)^2\biggr. \label{ham}
\end{eqnarray}
where ${\bf M}_i$ is the classical local magnetization normalized to $M=1$; $\psi_{i,\pm}$ are the $i$-th pair of Weyl fermion fields and ${\bf s}_{i,\pm}=\frac{1}2\psi^\dag_{i,\pm}\boldsymbol\sigma\psi_{i,\pm}$. $J>0$ is the ferromagnetic exchange amplitude; $K$ is Kondo coupling and $D$ is DM strength. The lattice constant is set to one. 

We focus on a single Weyl fermion field and back to the multi-Weyl fermion's later. Taking a flat metric on $T^3$ and  identifying ${\bf A}=\frac{K}2{\bf  M}$, the Hamiltonian density can be written as 
\begin{eqnarray}
-i\psi^\dag \boldsymbol\sigma\cdot(\nabla+i{\bf A})\psi+\frac{D}{K^2}\epsilon_{abc}A_a\partial_b A_c +\frac{J}{K^2}F_{ab}F_{ab}~
\end{eqnarray}
with Coulomb gauge $\nabla\cdot{\bf A}=0$ and the constraint $A=\frac{|K|}2$. This is a Chern-Simons-Dirac-Maxwell functional, referring to  the Chern-Simons-Dirac functional  ({\ref{CSD}).  The ground state of this Hamiltonian is determined by variating with ${\bf M}$ and $\psi$,
\begin{eqnarray}
&&-i\boldsymbol\sigma\cdot(\nabla+i\frac{K}2{\bf M})\psi=0\nonumber
\\
&&{\bf s}+\frac{D}{2K}\nabla\times {\bf M}-\frac{J}K\nabla^2{\bf M}=0.\label{9}
\end{eqnarray}
 When $J=0$, these equations  are exactly the SW-Eqs for $D<0$ and the F-Eqs for $D>0$ on $T^3$ \cite{3T}. When $K=0$, the Weyl semimetal decouples to ${\bf M}$. The solutions are well-known: The Weyl fermion corresponds to a momentum space monopole, i.e, $\vec{\cal B}_{\bf k}=\nabla_k\times i\langle\psi_k|\nabla_k|\psi_k\rangle=\frac{\bf k}{k^3}$  where $\psi_k$ is the Fourier component of $\psi$ \cite{w1} while the magnetization is a chiral magnet \cite{nag}
 \begin{eqnarray}
 {\bf M}={\bf e}_1\cos({\bf k}_0\cdot{\bf r}+\varphi_0)+{\rm sgn}(D) {\bf e}_2\sin({\bf k}_0\cdot{\bf r}+\varphi_0)\label{10}
 \end{eqnarray}
 where ${\bf e}_1\times{\bf e}_2={\bf k}_0/|{\bf k}_0|$ with $|{\bf k}_0|=|D|/4J$.  This helical spin texture has been experimentally observed in the cubic B20 compound MnSi \cite{sky1,sky2}. As an external magnetic field is applied, a conical magnetic structure which is continuously connected to the chiral magnet is stable in most  region of temperature-magnetic field phase diagram and ${\bf k}_0$ is pinned by the external magnetic field while a skyrmion lattice was observed in certain region in the phase diagram \cite{sky1,sky2}. For $J=K=0$  $E_2$ group symmetric solutions on $R^3$ were discussed in \cite{sol2}. 
 
 There is a "trivial solution" of Eq. (\ref{9}), the MWFM  whose magnetization is of a ferromagnetic order, say, ${\bf M}=(0,0,1)$ and $\chi$ is different a local phase from a free Weyl fermion with a monopolar Berry phase \cite{w1}.

 \noindent{\it SW monopoles with Chiral magnets. } We present a type of new solutions of the SWMEs (\ref{9}) which were not found on $R^3$.  Because of the periodic boundary conditions imposed in $T^3$, all solutions gave in the literature on $R^3$  \cite{freund,freund1,sol1,sol2,sol3} do not work.  As the Weyl equation does not change, the ansatz (\ref{4}) and (\ref{5}) still hold while Eq. (\ref{6}) is replaced by Eq. (\ref{9}). With a rescaling  $\chi=(|K|/\sqrt{2|D|})\psi$, it is 
  \begin{eqnarray}
 {\bf B}_0=-\frac{D}{|D|}{\bf B}+\frac{2J}{|D|}\nabla^2{\bf A}. \label{11}
 \end{eqnarray}
 The conditions to solve Eqs. (\ref{4}), (\ref{5}) and (\ref{11}) seem to be very strict: Besides the periodic boundary condition, one requires $A=\frac{|K|}2$ be a constant as well as $\nabla\cdot{\bf A}=\nabla\cdot{\bf B}_0=0$. 
We find such a solution, which is similar to Eq. (\ref{10}). For example, let us try a periodic chiral magnet solution ${\bf M}=\frac{Q}{K}(0,\cos Qx,\sin Q x)$, i.e., ${\bf A}=\frac{Q}{ 2}(0,\cos Q x,\sin Q x)$, 
$
{\bf B}=-\frac{Q^2}{2}(0,\cos Q x,\sin Qx)=-Q{\bf A}. $
Eq. (\ref{11}) reads
\begin{eqnarray}
{\bf B}_0=(\frac{D Q}{|D|}-\frac{2J Q^2}{|D|}){\bf A}.\label{13}
\end{eqnarray}
We see that the constraints $\nabla\cdot{\bf A}= \nabla\cdot{\bf B}_0=0$ are satisfied. 
Substituting Eq. (\ref{13}) into Eq. (\ref{5}), one has
\begin{eqnarray}
{\bf A}={\rm sgn} (D -2J Q)\frac{|Q|}{|K|}{\bf A}
\end{eqnarray}
To be a solution, one requires $|Q|=|K|$, which is also a request of $A=|K|/2$, and 
\begin{eqnarray}
{\rm sgn}  (D-2J Q)=1.
\end{eqnarray}
Therefore, as shown in Fig. \ref{fig1}(a), the F-Eqs ($D>0$) are of one such solution when $Q=-|K|<-\frac{D}{2J}$ and of two solutions with $Q=\pm |K|$ when $|K|< \frac{D}{2J}$. There  is no solution for $Q>D/2J$. The SW-Eqs ($D<0$) have such a solution when $Q=-|K|<D/2J$. When $J=0$, we see that there is no such a regular solution for the SW-Eqs while the F-Eqs are of two solutions with $Q=\pm|K|$, as expected. The sign of $Q$ gives the chirality of the magnetization (See Fig.\ref{fig1}(b) and (c)). 

\begin{figure}
\begin{center}
\includegraphics[width=7.5cm]{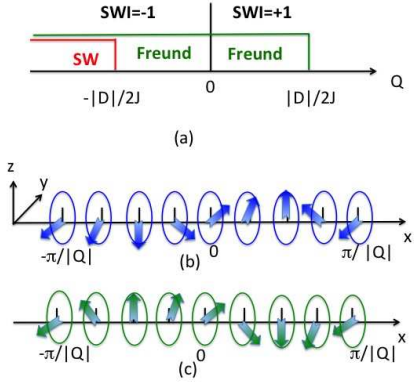}
\end{center}
 \caption{ \label{fig1} (Color online) (a) The solvable parameter region of the SW-Eqs and F-Eqs. (b) A right hand chiral magnet. (c) A left hand chiral magnet. } 
 \vspace{-4mm}
 \end{figure}

For a given chiral magnet, the Weyl fermion $\chi$ with $\chi^\dag\chi=B_0$  is given by
\begin{eqnarray}
\chi=\sqrt{\frac{Q^2(D-2JQ)(1+\sin Qx)}{4|D|}}\left(\begin{array}{cc}  \displaystyle1\\ \displaystyle \frac{i\cos Qx}{1+\sin Qx}\end{array}\right).
\end{eqnarray}
With an angle shift $Qx\to Qx+\varphi_0$, the chiral magnet is also a solution because it does not matter what is the initial angle of a chiral magnet. 

The chiral magnet ${\bf M}=\frac{Q}{K}(\sin (Qy+\varphi_0), 0,\cos(Qy+\varphi_0))$ in the same parameter region is also a solution. 

To see ${\bf A}=\frac{Q}{ 2}(\cos Q z,\sin Q z,0)$ is also a solution, we first consider a solution ${\bf A}'={\bf A}+(0,0,2/Q)$. The latter obeys Eqs. (\ref{5}) and (\ref{11}) but $A^2=\frac{Q^2}4+\frac 4{Q^2}$. However, ${\bf A}'$ differs a gauge transformation  from ${\bf A}$,.i.e., ${\bf A}'={\bf A}+\nabla f$ with $f=2z/Q$. Thus, we can conclude that general solutions for the SW-Eqs and F-Eqs, the SW monopole $(\chi, {\bf A})$ \cite{sol2},  are given by a chiral magnet 
\begin{eqnarray}
{\bf M}=\frac{|{\bf Q}|}{K}{\bf e}_1\sin({\bf Q}\cdot{\bf r}+\varphi_0)\pm{\bf e}_2\cos({\bf Q}\cdot{\bf r}+\varphi_0) \label{cmag}
\end{eqnarray}
where ${\bf e}_1\times{\bf  e}_2={\bf Q}/|{\bf Q}|$ and $|{\bf Q}|=|K|$ for $Q={\bf Q}\cdot{\bf r}/r$ that is restricted in the region shown in Fig. \ref{fig1}(a). We call $\chi$ the spheric Weyl fermion since it is massless when ${\bf Q}$ sweeps over a sphere $S^2$ with a radius $|K|$.  $\psi^\dag\psi=|D-2JQ|\ne 1$ implies the renormalization of the Weyl surface which resembles the renormalization of Fermi surface.

\begin{figure}
\begin{center}
\includegraphics[width=7.5cm]{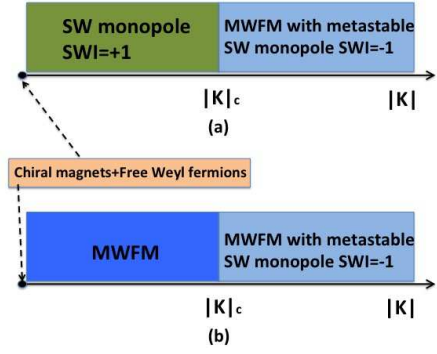}
\end{center}
 \caption{ \label{fig2} (Color online) (a) The phase diagram in $|K|$ for the F-Eqs. (b) The phase diagram in $|K|$ for the SW-Eqs. }
 \vspace{-4mm}
 \end{figure}

\noindent{\it Phase diagrams and Weyl metal. } While the Weyl fermions are massless the magnetization contribution to the energy of the SW monopole $(\chi,A_a)$ is given by
\begin{eqnarray}
E_Q=-\frac{1}2(DQ-2JK^2). \label{ge}
\end{eqnarray}
For the F-Eqs where $D>0$,  the energy of a solution with $Q=|K|<|K|_c=D/2J$ is less than zero and the SW monopoles are the ground states. It is a {\it Weyl metal} with a Wely fermion dispersion \cite{suppl} 
$$E_\psi(k)\propto \pm|Q|\sqrt k$$ 
for ${\bf k}\perp {\bf Q}$ (See the insertion in  Fig. \ref{fig3}(b)). When $Q=-|K|<0$, $E_Q=\frac{1}2(D|K|+2JK^2)>0$ and then the SW monopoles are metastable because the ground state is either the SW monopoles with $0<Q<|K|_c$ or the zero energy state, the MWFM when $Q<-|K|_c$. The phase diagram for the F-Eqs is shown in Fig. \ref{fig2}(a).  $K=0$ is an isolate point as discussed before (See Eq. (\ref{10})). 

For the SW-Eqs $(D<0)$, the SW monopoles exist only when $Q<-|D|/2J$ as shown in Fig. \ref{fig1}(a) with $E_Q=\frac{1}2(|DK|+2JK^2)>0$. Thus, they are metastable associated with the MWFM ground state. For $|Q|=|K|<|K|_c$, the system is in the MWFM phase without the metastable SW monopoles. Although the ground states of these two phases are the same the topological natures of them are different, which resembles arguments for a Weyl magnon topologically differing from a conventional antiferromagnetic order \cite{w7}. The phase diagrams are shown in Fig. \ref{fig2}(b).

\noindent{\it Seiberg-Witten invariants. } The SW monopoles are topological objects whose topological nature is portrayed by the SW invariants. Although a constant $\chi^\dag\chi$ means the SW monopole $(\chi, A_a)$ determined by Eq. (\ref{cmag})  is not square integrable, the moduli space may still compact  because $\chi$ is periodic \cite{hut}. 
The chiral magnets (\ref{cmag}) in the SW monopoles show that there are large degeneracy of the solutions of the SWMEs, i.e., the moduli space of the SWMEs are a two-dimensional sphere $S^2$  with a radius $|K|$ in the wave vector space, the {\it Weyl surface} of the Weyl metal \cite{note,suppl}.  For the SW-Eqs ($D<0$) with $|K|>|K|_c$, the moduli space consists of a single Weyl surface and the SW invariant is defined by the Euler number of the $S^2$ with its orientation, or by the winding number \cite{Aro,kita,suppl}
\begin{eqnarray}
SWI=\frac1{4\pi}\int_{S^2} ds_{Q_a}\epsilon_{abc}{\bf n}\cdot \partial_{Q_b}{\bf n}\times \partial_{Q_c}{\bf n}=-1\label{winding}
\end{eqnarray}
where  ${\bf n}=-{\bf Q}/|{\bf Q}|$ due to the chirality of the chiral magnets.  The metastable SW monopoles are of $SWI=-1$.  

For the F-Eqs, $SWI=0$ when $|K|<|K|_c$  because the number of the left chiral monopoles is the same as the number of the right ones. However, $E_Q>0$ for the SW monopoles with $Q<0$ while $E_Q<0$ for $Q>0$. Thus, $SWI=+1$ for the ground state.  
The metastable SW monopole for $Q<-|K|_c$ is of $SWI=-1$. 
We show these SW invariants in the phase diagrams (Fig. \ref{fig2}).

\noindent{\it External magnetic field. } We see that the ground states of the system are largely degenerate even $K$ is fixed.  It was known that an external magnetic field can pin the helical vector ${\bf k}_0$ in Eq. (\ref{10}) to the external field direction. The chiral magnets in the SW monopoles are
similar. Assuming the external field ${\bf h}=\nabla\times {\bf a}$ with $h$ being  a constant, the Hamiltonian reads
\begin{eqnarray}
{\cal H}&=&-i\chi^\dag \boldsymbol\sigma\cdot(\nabla+i{\bf a})\chi+K{\bf M}\cdot{\bf s}+ \frac{D}4{\bf M}\cdot \nabla\times{\bf M}\nonumber\\
&+&\frac{J}2(\nabla {\bf M})^2-{\bf h}\cdot{\bf M} 
\end{eqnarray}
For Freund equation in the SW monopole phase, the ferromagnetic order wins when $h>h_c=\frac{1}2|DQ-2JK^2|$.  When $h<h_c$, the magnetization part of the Hamiltonian is minimized by a conical spin structure \cite{sky1}, e.g., if ${\bf h}=(h,0,0)$, 
\begin{eqnarray}
{\bf M}=\frac{Q\sqrt{2-h^2}}{\sqrt{2}K}(\frac{h}{\sqrt{2-h^2}},\cos (Qx),\sin (Qx)).
\end{eqnarray}
 This means that the helical wave vector ${\bf Q}$ is oriented to parallel to the external field direction. This conical structure can continuously connect with the helical magnetic structure in a zero field.  The first order correction to the fermion part of the Hamiltonian has a vanishing mean field value. Therefore, this conical magnetization is stable and the SW monopole with the helical wave vector parallel to ${\bf h}$ is picked up from the large amount degeneracies. This single SW monopole means a zero-dimensional moduli space and its chirality defines $SWI=1$. For the SW-Eqs, a metastable $SWI=-1$ monopole is pinned by the external field.

\noindent{\it Multi-Weyl surfaces. } To see the mulit-Weyl surfaces' result, we consider a Weyl fermion field with velocity $-1$ in Eq. (\ref{ham}). Defining ${\bf M}'= -{\bf M}$, the SW monopole $(\psi,{\bf M}')$ obeys the same SWMEs as those for the Weyl fermion with the velocity $+1$. Thus, except a sign in the chiral magnet, the SW monopoles corresponding to different Weyl points are the same. Since there is an arbitrary angle $\varphi_0$, this sign is not important because they differ by an angle $\varphi_0=\pi$.  Interestingly, two SW monopoles for a pair of Weyl fermions with opposite velocities own the same SW invariant. 
The total SW invariant is the sum of the individual SW invariants with the same signs, comparing with a pair of  monopolar Weyl points whose "magnetic charges" are cancelled.

 \noindent{\it Fermi arcs and Fermi pockets at surfaces. }  A remarkable behavior of the Weyl semimetal is its exotic surface state: There are unclosed  Fermi arcs \cite{w1}.  In the MWFM phase, the Weyl fermions are semimetal and the Fermi arcs exist (Fig. \ref{fig3}(a)). In the SW monopole phase for F-Eqs,  one can show that the Berry curvature $\vec{\cal B}_k=0$.  This is consistent with the disappearance of the monopole in the momentum space. 
While, thus, there are no Fermi arcs, the surface states are metallic and there are Fermi pockets which are the projections of the Weyl surfaces. For instance, projecting to the surface for $z=$constant, the spheres $S^2$ are reduced to $S^1$ surrounding the Fermi pockets as shown in Fig. \ref{fig3}(b) if $|K|<k_w$, the separation between two Weyl points. When $|K|>k_w$, two Weyl spheres will connect and the topology of the Fermi pockets become more fruitful but that is not studied in details here. For the SW-Eqs, the surface states in ground states are the same of the free Weyl fermions. It is not studied here when the metastable SW monopoles are excited.

\begin{figure}
\begin{center}
\includegraphics[width=6.5cm]{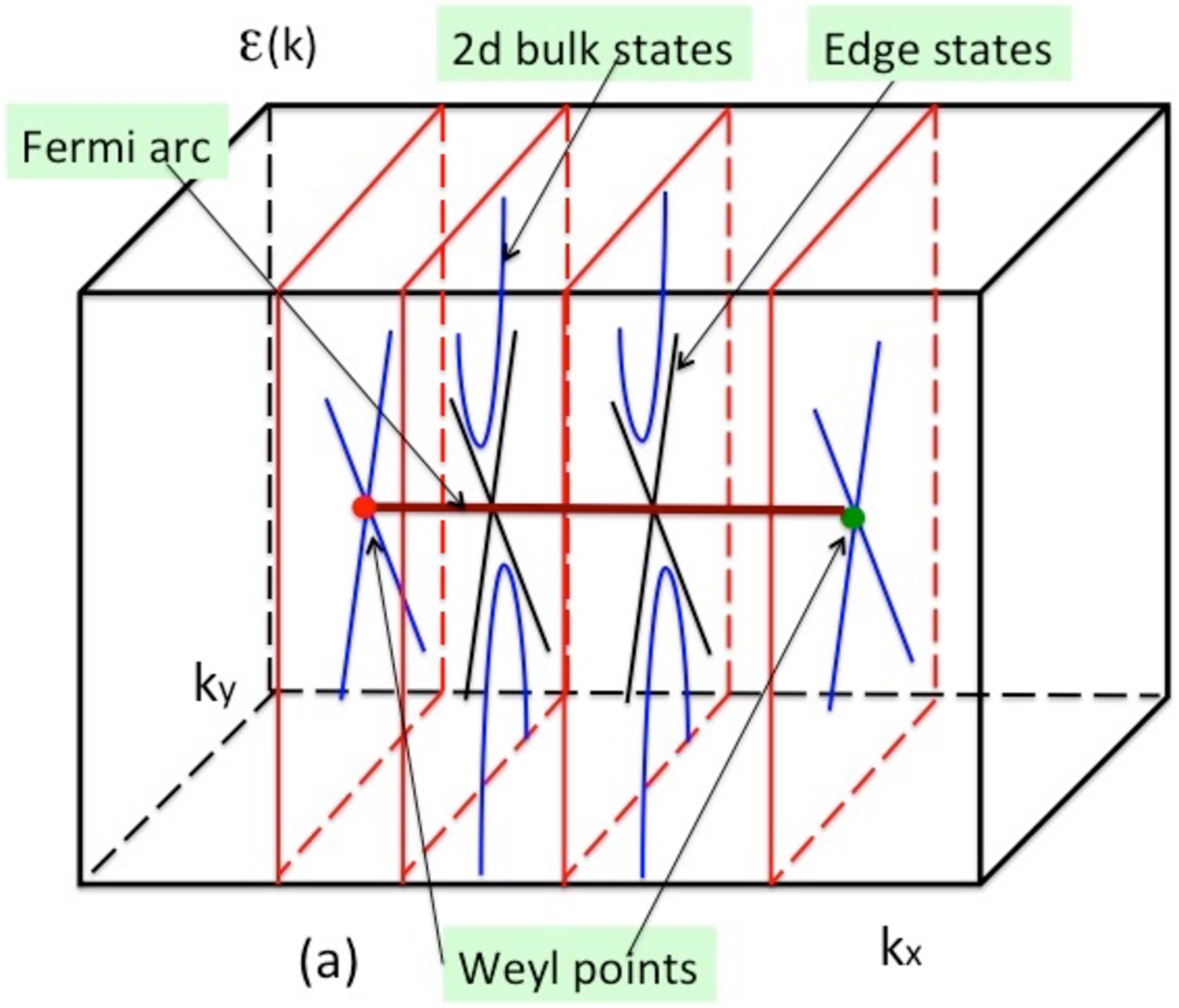}
\includegraphics[width=6.5cm]{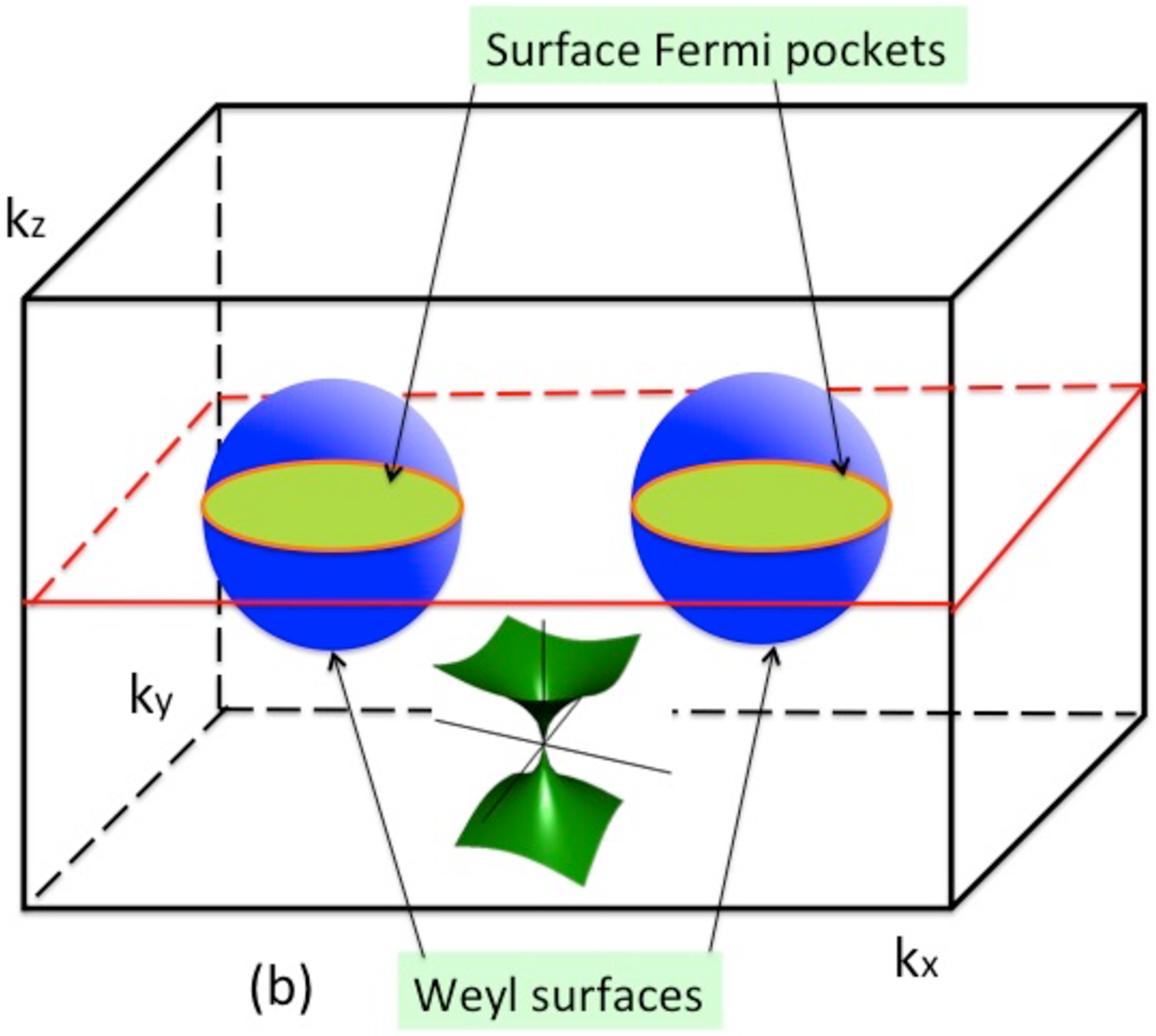}
\end{center}
 \caption{ \label{fig3} (Color online) (a) The Fermi arc for MWFM states. (b) The Weyl surfaces and surface Fermi pockets for SW monopoles. The insert is a sketch of $E_\psi$ at the point $(0,0,|Q|)$ on the Weyl surface.  }
 \vspace{-6mm}
\end{figure}

 \noindent{\it Possible real systems. } The ferromagnetically ordered
spinel compound HgCr$_2$Se$_4$ was proposed to be a
double Weyl semimetal with quadratic band touchings in the plane
normal to the ferromagnetic direction \cite{mater1,mater2}. If the double Weyl points are separated into two individual ones, the SW monopoles might emerge.  This material is also a candidate for a hybrid Weyl semimetal \cite{hws}. Very recently, two kinds of ferromagnetic and noncentrosymmetric topological semimetal materials have been discovered and  proposed: Sr$_{1-y}$Mn$_{1-z}$Sb$_2$ $(y, z < 0.10)$ \cite{mater3} and some  members of RAlX family of compounds (R=Ce, Pr, X=Si, Ge) \cite{mater4}. We  expect the SW monopoles can be possible topological objects of these systems.
 
\noindent{\it Conclusions. } We studied a physical model in which the Weyl (semi)metal couples to the local magnets. We found that the ground states of the system are determined by  the SWMEs. There are the "trivial" phase MWFM and non-trivial SW monopole phase (or the Weyl metal) as well as the MWFM with metastable SW monopoles. The critical points of the phase transitions  and the SWI on Weyl surfaces are determined.  Several further questions are listed in the {\it Supplemental Materials}.

 The author thanks L. Y. Hung, X. Luo, X. G. Wan and Y. -S. Wu for helpful discussions.  This work is supported by the 973 program of MOST of China (2012CB821402), NNSF of China (11174298, 11474061).

 \newpage
 
\centerline{\bf  Supplemental Materials }

 \subsection{Briefing of Seiberg-Witten monopoles on three dimensions}
  
 The Seiberg-Witten theory was originally defined on the four-dimensional space and the solutions of the Seiberg-Witten equations are named as the Seiberg-Witten monopole \cite{SW}. Witten found that the Donaldson invariants in $SU(2)$ anti-self dual Yang-Mills moduli space \cite{ds2} can be calculated by using the Seiberg-Witten invariant \cite{witten}.  Donaldson pointed out that the Seiberg-Witten invariant can also be defined in a three dimensional space \cite{donaldson}. Here we give a brief introduction to the Seiberg-Witten monopole equations and invariant in the three dimensions. Without concerning the supersymmetric Yang-Mills theory \cite{SW}, the Seiberg-Witten monopole equations can be thought of as to minimize the Chern-Simons-Dirac functional on $X$ \cite{donaldson} which reads
 \begin{eqnarray}
\int d^3{\bf r}\sqrt g[ -i\chi^\dag \boldsymbol\sigma\cdot(\nabla+i{\bf A}+i\boldsymbol\omega)\chi\pm\epsilon_{abc}A_a\partial_b A_c]~\label{CSD}
\end{eqnarray}
where $\chi^\dag=(\alpha^*,\beta^*)$ is a Weyl spinor; $A_a$ is a $U(1)$ gauge field; $\boldsymbol\sigma$ are Pauli matrices. $g=\det g_{ab}$ with the metric $g_{ab}$ on $X$ and $\boldsymbol\omega$ is the spin connection. The repeat indices  imply summation over $a,b,c=1,2,3$.  By variating with $\chi^\dag$ and $A_a$, one has the Seiberg-Witten monopole equationss on $X$
\begin{eqnarray}
&&\boldsymbol\sigma\cdot(\nabla+i{\bf A}+i\boldsymbol\omega )\chi=0\label{sw1}\\
&&\chi^\dag\sigma^a\chi=\pm\epsilon_{abc}\partial_bA_c. \label{sw2}
\end{eqnarray}
The equations with the plus sign in (\ref{sw2}) are called Seiberg-Witten equations while the minus sign corresponding to Freund equations \cite{freund,freund1}.  The solution of Seiberg-Witten monopole equations, a pair of $(A_a,\chi)$,  is called {\it Seiberg-Witten monopole}, in the sense that $U(1)$ as the maximal abelian subgroup is dual to  $SU(2)$ gauge group  and the "monopole" $\chi$ is dual to  the "electric charge" in the original $SU(2)$ supersymmetry Yang-Mill  theory \cite{SW,witten}. The square integrable solution space of the Seiberg-Witten monopole equations is named as the moduli space of the Seiberg-Witten  monopoles. The Seiberg-Witten  invariant is basically defined by the Euler number of the moduli space  with its orientation\cite{witten,donaldson,bohn}. In certain cases, the moduli spaces are zero-dimensional, i.e., discrete solutions. One can assign a sign to each discrete solution and the sum of them defines the Seiberg-Witten  invariant \cite{witten,donaldson}. This invariant is equivalent to Casson invariant \cite{fl} according to the Seiberg-Witten-Floer cohomology \cite{donaldson,bohn}. On a three-dimensional  torus $T^3$,  one can show that a periodic Seiberg-Witten-Floer theory also has a compact moduli space \cite{hut}.

\subsection { Ansatz on a flat space }

 On the euclidean space $R^3$, the metric is flat and one can take $\boldsymbol\omega=0$. To solve Eqs. (\ref{sw1}) and (\ref{sw2}) in $R^3$, one can transfer the Seiberg-Witten monopole equations to an ansatz \cite{sol1,sol2,sol3}
\begin{eqnarray}
&&\chi=\frac{1}{\sqrt{2(B_0+B_{03})}}\left(\begin{array}{c} B_0+B_{03}\\B_{01}+iB_{02}\end{array}\right) \label{4}\\
&&{\bf A}=-\frac{\nabla\times {\bf B}_0}{2B_0}-\frac{B_{01}\nabla B_{02}-B_{02}\nabla B_{01}}{2B_0(B_0+B_{03})} \label{5} \\
&&{\bf B}_0=\pm{\bf B}=\pm\nabla\times {\bf A} \label{6}
\end{eqnarray} 
with ${\bf B}_0=\chi^\dag\boldsymbol\sigma\chi$  obeying $\nabla\cdot {\bf B}_0=0$. One can check $\chi^\dag\chi=B_0$.  Many solutions of the Seiberg-Witten monopole equations in  $R^3$ were known \cite{freund,freund1,sol1,sol2,sol3}. Since $R^3$ is flat,  there are only singular solutions for the Seiberg-Witten equations and no square integrable ones are allowed \cite{witten} while both singular and non-singular solutions of the Freund equations exist. The Seiberg-Witten invariant is then of no definition for the Seiberg-Witten equations on $R^3$ while it is always trivial for the Freund equations because the triviality of $R^3$. The ansatz (\ref{4}) and (\ref{5}) also hold on $T^3$ with periodic boundary conditions.

\subsection{Seiberg-Witten invariant and winding number}

The energy of our Seiberg-Witten  monopole solution is given by Eq. (17) in the main text. For $D<0$ or $D>0$, 
\begin{eqnarray}
E_q=-\frac{1}2(Dq-2Jq^2)=\pm\frac{1}2|D|q+Jq^2
\end{eqnarray}
where $q=\pm |K|$.
For a given $D$, say, $D<0$,  this gives a two level system because $q$ may  either be positive or negative. 
 The projective operator $P=\frac{1}2(1+{\bf n}\cdot \boldsymbol\tau)$, where $\boldsymbol\tau$ are the Pauli matrices, projects a state to the lower level. The Chern number of the lower level is then  defined as \cite{Aro, kita}
 \begin{eqnarray}
 c_1=\frac{i}{4\pi}\int _{S^2}{\rm Tr}(dP\wedge PdP)=\frac{1}{2\pi i}\int d^2s_q\epsilon_{ab}F^{ab},
 \end{eqnarray}
 where
 $$ F_{ab}={\rm Tr}\biggl[P\biggl(\frac{\partial P}{\partial{q_a}} \frac{\partial P}{\partial{q_b}}-\frac{\partial P}{\partial{q_b}} \frac{\partial P}{\partial{q_a}}\biggr)\biggr]$$ 
 Eq. (8) exactly gives the winding number (18) in the main text with ${\bf n}=\frac{\bf q}{|\bf q|}$, i.e., $SWI=1$. For $D<0$, the project operator is $P=\frac{1}2(1+{\bf n}\cdot \boldsymbol\tau)$  with ${\bf n}=-\frac{\bf q}{|\bf q|}$and then the winding number is $SWI=-1$, the result of Eq. (18) in the main text.

 \subsection{Gapless Weyl fermions}
The Dirac equations are given by 
\begin{eqnarray}
-i\sigma^a(\partial_a+iA_a)\chi=E_\psi\chi.\nonumber
\end{eqnarray}
Acting $-i\sigma^a(\partial_a+iA_a)$ on the both sides of the Dirac equations, one gets
\begin{eqnarray}
-(\nabla^2+2i{\bf A}\cdot\nabla-{\bf A}^2+i\nabla\cdot{\bf A}+{\bf B}\cdot\boldsymbol\sigma)\chi=E_\psi^2\chi. \nonumber
\end{eqnarray}
Without losing of the generality, considering ${\bf A}=\frac{Q}2(0,\cos Qx,\sin Qx)$ and defining $\chi=e^{ik_yy+ik_zz}\chi(x)$,  we have
 \begin{eqnarray}
&&(-\partial_x^2+Q\cos Qx k_y+Q\sin Qx k_z-Q^2/4)\chi(x)\nonumber\\
&&-Q^2/2(\cos Qx\sigma^y+\sin Qx \sigma^z)\chi(x)\nonumber\\
&&=(E_\psi^2-k^2)\chi(x)\nonumber
\end{eqnarray}
where $k^2=k_y^2+k_z^2$.  Define $\chi(x)=G(x,k)\chi_0(x)$ where $\chi_0$ is the solution when $k=0$ and $E_\psi=0$, i.e., Eq. (15) in the main text, and $G(x,k))$ is a $2\times 2$ matrix with $G(x,0)=1$. In the long wave length limit, $\chi(x)=G(x,k)\chi_0=(1+G_1k+G_2k^2+O(k^3))\chi_0=\chi_0+k\chi_1+k^2\chi_2+O(k^3)$, $E_\psi^2=k|Q|\varepsilon_1^2+k^2 \varepsilon_2^2+O(k^3)$, then
\begin{eqnarray}
&&(-\partial_x^2+Q\cos Qx k_y+Q\sin Qx k_z-Q^2/4)\nonumber\\
&&\times(1+G_1 k+G_2k^2+\cdots)\chi_0-Q^2/2(\cos Qx\sigma^y+\sin Qx \sigma^z)\nonumber\\
&&\times(1+G_1 k+G_2k^2+\cdots)\chi_0\nonumber\\
&&=(k\varepsilon_1^2+k^2 \varepsilon_2^2-k^2+\cdots)(1+G_1 k+G_2k^2+\cdots)\chi_0\nonumber
\end{eqnarray} 

The zero order equation is just Eq. (15) in the main text. The first order equation reads
\begin{eqnarray}
&&[-\partial_x^2-\frac{Q^2}4-\frac{Q^2}2(\cos Qx\sigma^y+\sin Qx \sigma^z)]G_1\chi_0\nonumber\\
&&+Q(\cos Qx \cos\theta+\sin Qx \sin \theta-\varepsilon_1^2)\chi_0=0,\nonumber
\end{eqnarray} 
where $k_y=k\cos\theta$ and $k_z=k\sin\theta$. The solution of this equations determines the dispersion of the Weyl fermions
\begin{eqnarray}
E_\psi=\pm\sqrt {|Q|}|\varepsilon_1|\sqrt k+O(k).
\end{eqnarray}

\subsection{Questions}

Finally, we list the following further questions can be asked  in order:
(i) How does the giant negative magnetoresistance change its magnitude? 
(ii)What are finite temperature behaviors of the whole system?  Is there a phase in which the Weyl fermions couple to skyrmions?   (iii) Instead of the classical magnetization, how a quantum spin fluctuate the SW monopoles, the phase diagrams and the dynamics of the system? (iv) Are there two-dimensional physical models which associate with the two-dimensional SW monopole equations?


\begin{thebibliography}{99}
\bibitem{note} The angle $\varphi_0$ arises $S^1$-degeneracies but a separated $S^1$ does not contribute to the SW invariants.
\end{thebibliography}
\end{document}